\documentclass[onecolumn,nofootinbib,showpacs,preprintnumbers,amsmath,amssymb]{revtex4-1}

\usepackage{amsfonts}
\usepackage{amsmath}
\usepackage{amssymb}
\usepackage{amstext}
\usepackage{amsthm}
\usepackage{graphicx}
\usepackage{dcolumn}
\usepackage{bm}
\usepackage{color}
\usepackage{float}

\begin{document}

\title{A GDP-driven model for the binary and weighted structure\\ of the International Trade Network }

\author{Assaf Almog}
\affiliation{Instituut-Lorentz for Theoretical Physics,Leiden Institute of Physics, University of Leiden, Niels Bohrweg 2, 2333 CA Leiden (The Netherlands)}

\author{Tiziano Squartini}
\affiliation{“Sapienza” University of Rome, P.le Aldo Moro 5, 00185 Rome (Italy)}
\affiliation{Instituut-Lorentz for Theoretical Physics, Leiden Institute of Physics, University of Leiden, Niels Bohrweg 2, 2333 CA Leiden (The Netherlands)}

\author{Diego Garlaschelli}
\affiliation{Instituut-Lorentz for Theoretical Physics, Leiden Institute of Physics, University of Leiden, Niels Bohrweg 2, 2333 CA Leiden (The Netherlands)}

\date{\today}

\begin{abstract}
Recent events such as the global financial crisis have renewed the interest in the topic of economic networks. 
One of the main channels of shock propagation among countries is the International Trade Network (ITN). 
Two important models for the ITN structure,
the classical gravity model of trade (more popular among economists) and the fitness model (more popular among networks scientists), are both limited to the characterization of only one representation of the ITN. 
The gravity model satisfactorily predicts the volume of trade between connected countries, but cannot reproduce the observed missing links (i.e. the topology). On the other hand, the fitness model can successfully replicate the topology of the ITN, but cannot predict the volumes. 
This paper tries to make an important step forward in the unification of those two frameworks, by proposing a new  GDP-driven model which can simultaneously reproduce the binary and the weighted properties of the ITN.
Specifically, we adopt a maximum-entropy approach where both the degree and the strength of each node is preserved. We then identify strong nonlinear relationships between the GDP and the parameters of the model.
This ultimately results in a weighted generalization of the fitness model of trade, where the GDP plays the role of a `macroeconomic fitness' shaping the binary and the weighted structure of the ITN simultaneously. 
Our model mathematically highlights an important asymmetry in the role of binary and weighted network properties, namely the fact that binary properties can be inferred without the knowledge of weighted ones, while the opposite is not true. 
\end{abstract}

\maketitle

\section{Introduction}

After the 2008 financial crisis, it has become clear that a better understanding of the mechanisms and dynamics underlying the networked worldwide economy is vital \cite{economicnetworks}. 
Among the possible channels of interaction among countries, international trade plays a major role \cite{Kali1,Kali2,integration}.
Combined together, the worldwide trade relations can be interpreted as the connections of a complex network, the International Trade Network (ITN) \cite{Serrano,Diego1,Garlaschelli,Vespignani,Caldarelli,Fagiolo1,Fagiolo2,Fagiolo3,Barigozzi,DeBene1,Squartini1,Squartini2,Squartini3,Fronczak,Squartini7,Pietronero}, whose understanding and modeling is one of the traditional goals of macroeconomics.
The standard model of non-zero trade flows, inferring the volume of bilateral trade between any two countries from the knowledge of their Gross Domestic Product ($GDP$) and mutual geographic distance ($D$), is the so-called `gravity model' of trade \cite{Tinbergen1, Tinbergen2, Giorgio_gravity0,Giorgio_gravity,DeBene2}.
In its simplest form, the gravity model predicts that the volume of trade between countries $i$ and $j$ is
\begin{equation}
F_{ij}= \alpha\frac{GDP^{\beta}_i \cdot GDP^{\beta}_j}{D^{\gamma}_{ij}}.
\label{eq_gravity}
\end{equation}
Note that the above expression, as well as the rest of this paper, focuses on the undirected version of the network for simplicity. In this representation, the trade from country $i$ to country $j$ and the trade rom country $j$ to country $i$ are combined together. 
Given the highly symmetric structure of the ITN, this simplification retains all the basic network properties of the system \cite{Garlaschelli,Squartini1,Squartini2,myreciprocity,Giorgio_symmetry,mywreciprocity}.

In its simplest form, the gravity model is fitted on the non-zero weights observed between all pairs of \emph{connected} countries. This means that the model can predict the pair-specific volume of trade only {\it after} the presence of the trade relation itself has been established \cite{Giorgio_gravity}. This intrinsic limitation is alarming, since almost half of the links in the ITN are missing \cite{Diego1,Squartini1,Squartini2,Squartini3}. 
Although several improvements and generalizations of the standard gravity model have been proposed to overcome this problem (see \cite{Giorgio_gravity,DeBene2} for excellent reviews), so far none of them succeeded in reproducing the observed complex topology and the observed volumes simultaneously.
Moreover, the various attempts have not been conceived under a unique theoretical framework and are therefore based on the combination of different mechanisms (e.g. one for establishing the presence of a trade relation, and one for establishing its intensity). 
In general, the challenge of successfully predicting, via only one mechanism, both trade probabilities and trade volumes remains an open problem.

Over the past years, the problem of replicating the observed structure of the ITN has been extensively approached using network models  \cite{Diego1,Garlaschelli,Caldarelli,Fronczak}
and, more indirectly, maximum-entropy techniques to reconstruct networks from partial information \cite{Squartini1,Squartini2,Squartini3,Wells,Bargigli,Musmeci,Caldarelli 2}.
These studies have focused both on the purely binary architecture (defined solely by the existence of trade exchanges between world countries) \cite{Diego1,Garlaschelli,Caldarelli,Squartini1,Musmeci} and on the weighted structure (when also the magnitude of these interactions is taken into account) \cite{Squartini2,Squartini3,Fronczak}. 
What clearly emerges is that both topological and weighted properties of the network are deeply connected with purely macroeconomic properties (in particular the GDP) governing bilateral trade volumes \cite{Kali1,Kali2,Serrano,Diego1,Garlaschelli,Caldarelli,Fagiolo1,Fagiolo2,Fagiolo3,Barigozzi,DeBene1,Fronczak}.
However, it has also been clarified that, while the knowledge of the degree sequence (i.e. the number of trade partners for each country) allows to infer the the entire binary structure of the network with great accuracy \cite{Squartini1,Squartini3}, the knowledge of the strength sequence (i.e. the total volume of trade for each country) gives a very poor prediction of all network properties \cite{Squartini2,Squartini3}.
Indeed, the network inferred only from the strength sequence has a trivial topology, being much denser (if integer link weights are assumed \cite{Squartini2}) or even fully connected (if continuous weights are assumed \cite{Fronczak}), and in any case much more homogeneous than the empirical one.
This limitation leads back to the main drawback of the gravity model.
Indeed, it has been shown that a simplified version of the gravity model (with $\beta=1$ and $\gamma=0$) can be recovered as a particular case of a maximum-entropy model with given strength sequence (and continuous link weights) \cite{Fronczak}.

Combined together, the high informativeness of the degree sequence for the binary representation of the ITN and the low informativeness of the strength sequence for its weighted representation contradict the naive expectation that, once aggregated at the country level, weighted structural properties (the strengths) are \emph{per se} more informative than purely binary properties (the degrees).
This puzzle has generated further interest around the challenge of finding a unique mechanism predicting link probabilities and link weights simultaneously.

As a step forward in this direction, an improved reconstruction approach \cite{Squartini5}, based on an analytical maximum-likelihood estimation method \cite{MaxEntropy}, has been recently proposed  in order to construct more sophisticated maximum-entropy ensembles of weighted networks. 
This approach builds on previous mathematical results \cite{Garlaschelli2} characterizing a network ensemble where both the degree and the strength sequences are constrained. The graph probability is the so-called generalized Bose-Fermi distribution \cite{Garlaschelli2}, and the resulting network model goes under the name of Enhanced Configuration Model (ECM) \cite{Squartini5}.
When used to reconstruct the properties of several empirical networks, the ECM shows a significant improvement with respect to the case where either only the degree sequence (Binary Configuration Model, BCM for short) or only the strength sequence (Weighted Configuration Model, WCM for short) is constrained.
One therefore expects that combining the knowledge of strengths and degrees is precisely the ingredient required in order to successfully reproduce the ITN from purely local information.
Indeed, a more recent study has showed that, when applied to international trade data (both aggregated and commodity-specific), the method successfully reproduces the key properties of the ITN, across different years and for different levels of aggregation (i.e. for different commodity-specific layers) \cite{Squartini6}.

However, in itself the ECM is a network reconstruction method, rather than a genuine model of network formation. To turn it into a proper network model for the ITN structure, it would be necessary to find a macroeconomic interpretation for the underlying variables involved in the method. 
This operation would correspond to what has already been separately performed at a purely binary level (by identifying a strong relationship between the GDP and the variable controlling the degree of a country in the BCM \cite{Diego1,Diego2}) and at a purely weighted level (by finding a relationship between the GDP and the variable controlling the strength of a country in the WCM \cite{Fronczak}, in the same spirit of the gravity model).
Generalizing the above results to the combination of strengths and degrees is not obvious, given the different mathematical expressions characterizing the ECM, the BCM and the WCM. 

In this paper we show that, indeed, the variables of the ECM are all strongly correlated with the GDP. This result gives a macroeconomic interpretation of the parameters' values satisfying the ITN constraints.
Reversing the perspective, this result enables us to introduce the first GDP-driven model that successfully reproduces the binary and the weighted properties of the ITN simultaneously.
Finally, we show that the ECM model can be replaced by a simpler, two-step model that reconciles the binary projection of the ECM model with the topology predicted by the BCM.
These results represent a promising step forward in the formulation of a unified model for the structure of the ITN.

\begin{figure*}{}
\includegraphics[scale=0.6]{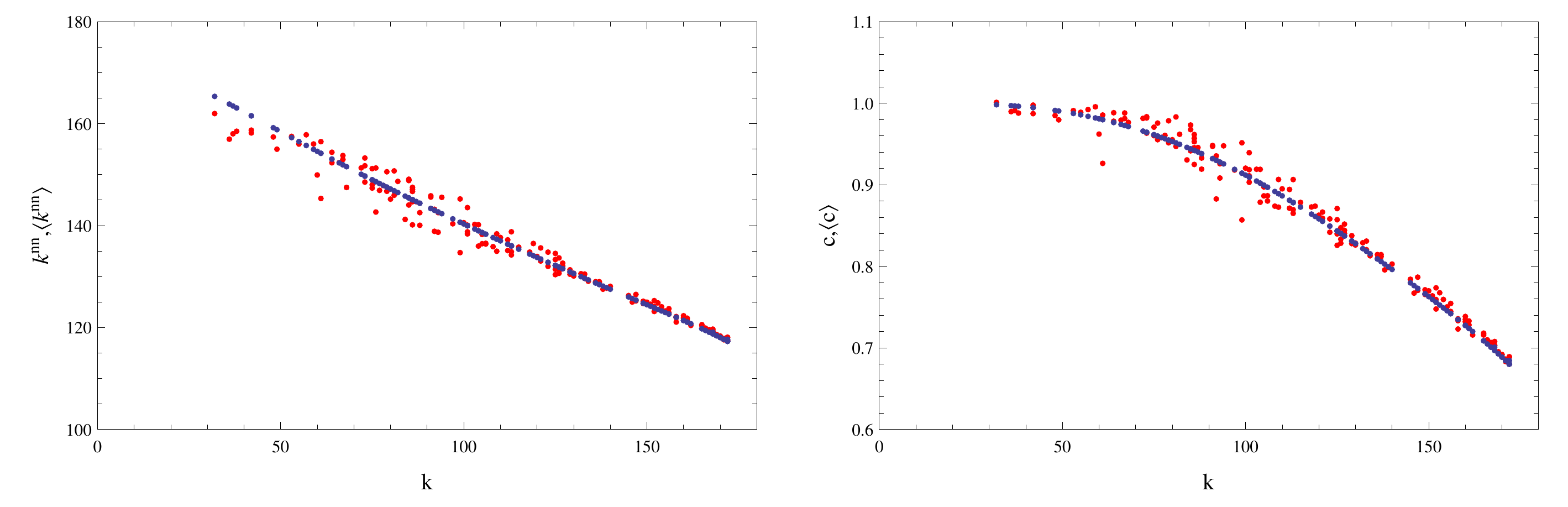}
\caption{\label{fig:epsart} Comparison between the observed undirected binary
properties (red points) and the corresponding ensemble averages of the BCM (blue points) for the aggregated ITN in the 2002 snapshot.
Left panel: Average Nearest Neighbor Degree $k^{nn}$ versus degree $k_i$.
Right panel: Binary Clustering Coefficient $C_i$ versus degree $k_i$.
}
\label{fig1}
\end{figure*}

\section{Maximum-entropy approaches to the ITN}


Since our results are a generalization of previous maximum-entropy approaches to the characterization of the ITN, in this section we first briefly review the main results of those approaches, while our new findings are presented in the next section.
In so doing, we gradually introduce the mathematical building blocks of our analysis and illustrate our main motivations.
Moreover, since previous studies have used different data sets, we also recalculate the quantities of interest on the same data set that we will use later for our own investigation.
This allows us to align the results of previous approaches and properly compare them with our new findings.

\subsection{Data}

We use yearly bilateral data on exports and imports from the United Nations Commodity Trade Database (UN COMTRADE) \cite{comtrade} from year 1992 to 2004.
The sample refers to 13 years, 1992-2004, represented in current U.S. dollars, and disaggregated over 97 commodity classes.
In this paper we analyze the aggregated level, which results in 13 yearly temporal snapshots of  undirected total trade flows.
Our network consists of $N = 173$ countries, present in the data throughout the considered temporal interval.\\

This data set was the subject of many studies exploring both purely  the binary representation, and its full weighted representation \cite{Squartini1,Squartini2,Squartini5,Squartini6}. 
Another data set which is widely used to represent the ITN network is the trade data collected by Gleditsch \cite{Gleditsch}.
The data contain the detailed list of bilateral import and export volumes, for each country in the period 1950-2000.

\subsection{Binary structure}

If one focuses solely on the binary undirected projection of the ITN, then the Binary Configuration Model (BCM) represents a very successful maximum-entropy model. 
In the BCM, the local knowledge of the number of trade partners of each country, i.e. the degree sequence, is specified.
It has been shown that higher-order properties of the ITN can be simply traced back to the knowledge of the degree sequence \cite{Squartini1}.
This result adds considerable information to the standard results of traditional macroeconomic analyses of international trade. 
In particular, it suggests that the degree sequence, which is a purely topological property, needs to be considered as an important target quantity that international trade models, in contrast with the mainstream approaches in economics,  should aim at reproducing \cite{Squartini1,Squartini3}.

Let us first represent the observed structure of the ITN as a weighted undirected network specified by the square matrix $\mathbf{W}^*$, where  the specific entry $w^*_{ij}$ represents the weight of the link between country $i$ and country $j$. 
Then, let us represent the binary projection of the network in terms of the binary adjacency matrix $\mathbf{A}^*$, with entries defined as $a_{ij}^*= \Theta[w^*_{ij}],\:\forall\:{i,j}$.
A maximum-entropy ensemble of networks is a collection of graphs where each graph is assigned a probability of occurrence determined by the choice of some constraints.
The BCM is a maximum-entropy ensemble of binary graphs, each denoted by a generic matrix $\mathbf{A}$, where the chosen constraint is the degree sequence. 
In the canonical formalism  \cite{MaxEntropy}, the latter can be constrained by writing the following Hamiltonian:
\begin{equation}
H(\mathbf{A})=\sum_{i=1}^N \theta_i k_i (\mathbf{A})
\end{equation}
where the degree sequence  is defined as $k_i(\mathbf{A})=\sum_{j\neq i}^N a_{ij}=\sum_{j\neq i}^N \Theta[w_{ij}],\:\forall\:{i}$.
As a result of the constrained maximization of the entropy \cite{MaxEntropy}, the probability of a given configuration $\mathbf{A}$ can be written as 
\begin{equation}
P(\mathbf{A})=
\frac{e^{-H(\mathbf{A})}}{\sum_{\mathbf{A}'}e^{-H(\mathbf{A}')}}=\prod_{i<j} \left[ \frac{z_iz_j}{1+z_iz_j}\right]^{a_{ij}}\left[ \frac{1}{1+z_iz_j}\right]^{1-a_{ij}}
\end{equation}
where $z_i \equiv e^{-\theta_i}$ and $p_{ij} \equiv   \frac{z_i z_j}{1+z_iz_j}$.
The latter represents the probability of forming a link between nodes $i$ and $j$, which is also the expected value  
\begin{equation}
\langle a_{ij}\rangle= \frac{z_i z_j}{1+z_iz_j}=p_{ij}.
\label{eq_pzz}
\end{equation}  
According to the maximum-likelihood method proposed in \cite{MaxEntropy}, the vector of unknowns $\vec{z}$ can be numerically found by solving the system of $N$ coupled equations
\begin{equation}
\langle k_i\rangle=\sum_{j\neq i}^N p_{ij}=k_i(\mathbf{A^*})\quad\forall i
\label{MaxLike}
\end{equation}
where the expected value of each degree $k_i$ is matched to the observed value $k_i(\mathbf{A^*})$ in the real network $\mathbf{A^*}$.
The (unique) solution will be indicated as $\vec{z}^*$. 
When inserted back into eq.(\ref{eq_pzz}), this solution allows to analytically describe the binary ensemble matching the observed constraints. Being the result of the maximization of the entropy, this ensemble represents the least biased estimate of the network structure, based only on the knowledge of the empirical degree sequence.

In Fig. \ref{fig1} we plot some higher-order topological properties of the ITN as a function of the degree of nodes, for the 2002 snapshot. 
These properties are the so-called average nearest neighbor degree and the clustering coefficient. 
For both quantities, we plot
the observed values (red points) and the corresponding expected values predicted by the BCM (blue points).
The exact expressions for both  empirical and expected quantities are provided in the Appendix.
We see that the expected values are in very close agreement with the observed properties. These results replicate recent findings \cite{Squartini1,Squartini3} based on the same UN COMTRADE data. 
They show that at a binary level, the degree correlations (disassortativity) and clustering structure of the ITN are excellently reproduced by
the BCM. 
As we also confirmed in the present analysis, these results were found to be very robust, as they hold true over time and for various resolutions (i.e., for different levels of aggregation of traded commodities) \cite{Squartini1,Squartini3}.

\subsubsection{Relation with the Fitness Model}
\begin{figure}{}
\includegraphics[scale=0.7]{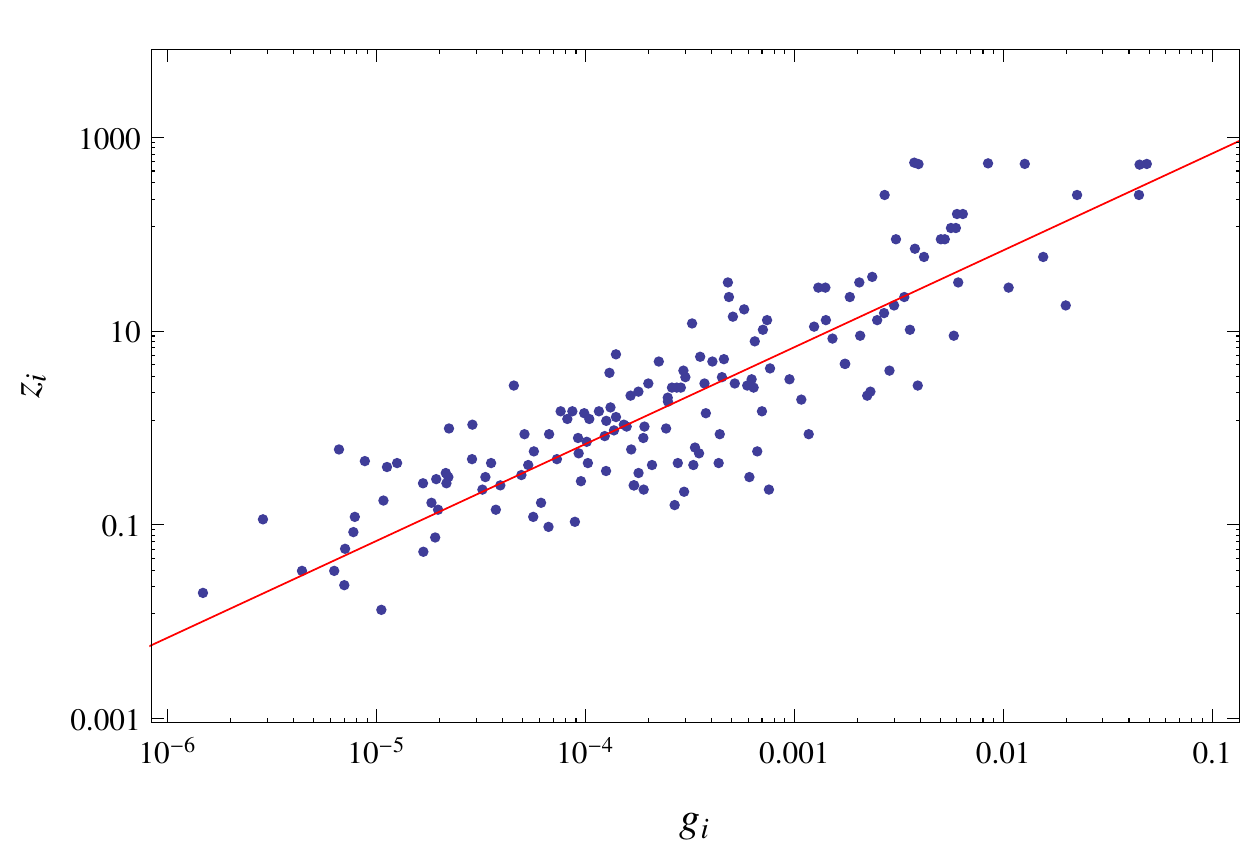}
\caption{\label{fig:epsart} The calculated $z_i$, compared with  the $g_i$ (re-scaled GDP) for each country for the undirected binary aggregated ITN in the 2002 snapshot,
with the linear fit $z_i=\sqrt{a}\cdot g_i$ (red line).}
\label{fig2}
\end{figure}

It should be noted that eq.(\ref{eq_pzz}) can be thought of as a particular case of the so-called \emph{Fitness Model} \cite{fitness}, which is a popular model of binary networks where the connection probability $p_{ij}$ is assumed to be a function of the values of some `fitness' characterizing each vertex. 
Indeed, the variables $\vec{z}^*$ can be treated as fitness parameters \cite{Diego1,Diego2} which control the probability of forming a link.
A very interesting correlation between a fitness parameter of a country (assigned by the model) and the $GDP$ of the same country was found \cite{Diego2}. This relation is replicated here in Fig. \ref{fig2}, where the rescaled $GDP$ of each country  ($g_i \equiv \frac{GDP_i}{\sum_i GDP_i}$) is compared to the  value of the fitness parameter $z_i^*$ obtained by solving eq.(\ref{MaxLike}). The red line is a linear fit of the type
\begin{equation}
z_i=\sqrt{a}\cdot g_i
\label{eq_zz}.
\end{equation}

This leads to a more economic interpretation where the fitness parameters can be replaced (up to a proportionality constant) with the $GDP$ of countries, and used to reproduce the  properties of the network. 
This procedure, first adopted in \cite{Diego1}, can give predictions for the network based only on macroeconomic properties of countries, and reveals the importance of the $GDP$ in to the binary structure of the ITN. 
Importantly, this observation was the first empirical evidence in favour of the fitness model as a powerful network model \cite{fitness}.
Likewise, other studies have shown that the observed topological properties turn out
to be important in explaining macroeconomics dynamics \cite{Kali1,Kali2}. 

\begin{figure*}{}
\includegraphics[scale=0.6]{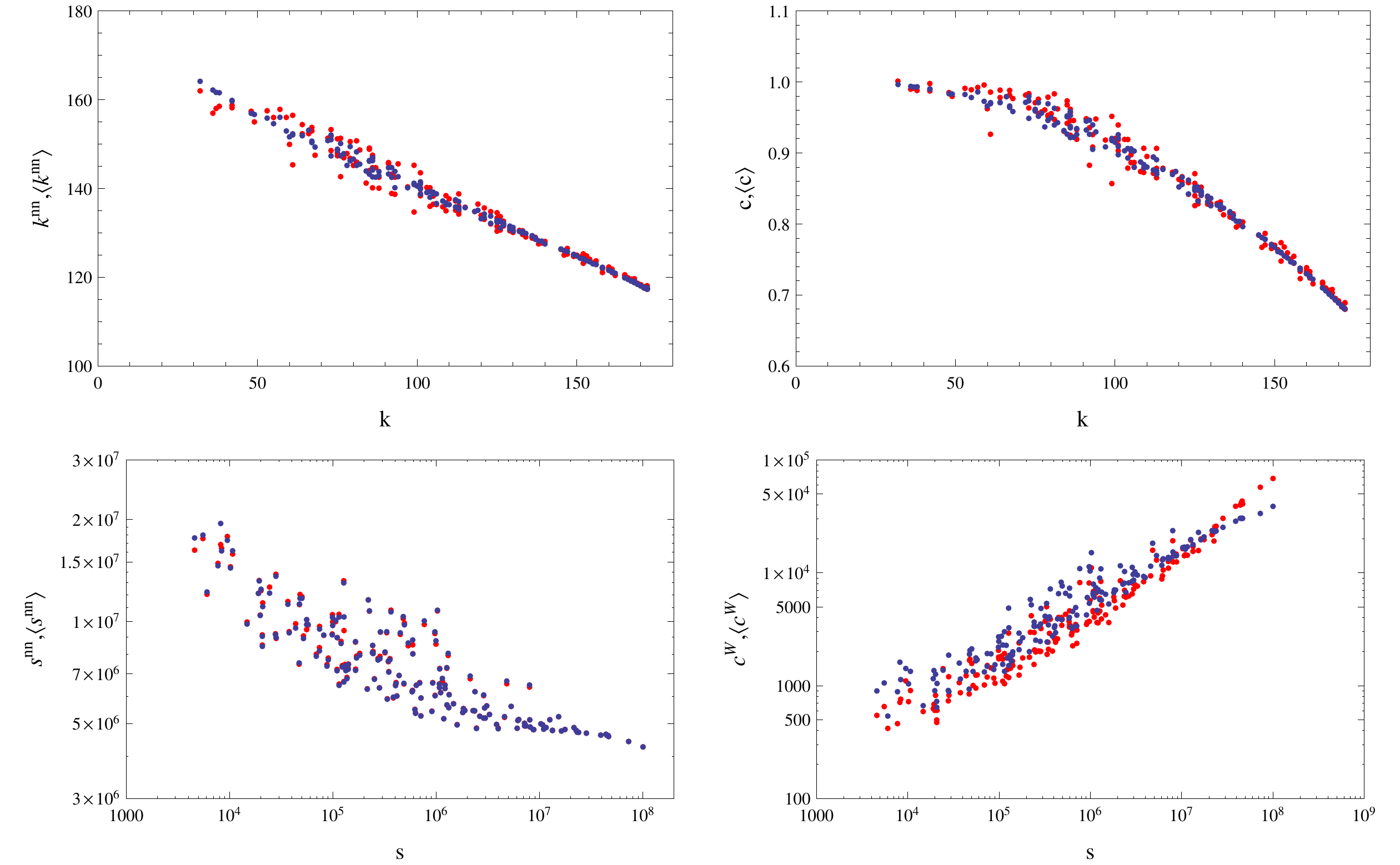}
\caption{\label{fig:epsart}Comparison between the observed undirected binary and weighted properties (red points) and the corresponding ensemble averages of the the ECM
(blue points) for the aggregated ITN in the 2002 snapshot.
Top left panel: Average Nearest Neighbor Degree $k^{nn}$ versus degree $k_i$;
Top right panel: Binary Clustering Coefficient $C_i$ versus degree $k_i$ ; 
Bottom left panel: Average Nearest Neighbor Strength $s^{nn}$ versus strength $s_i$ ;
Bottom right panel: Weighted Clustering Coefficient $C^{W}$ versus strength $s_i$ .
}
\label{fig3}
\end{figure*}

\subsection{Weighted structure}

Despite the importance of the topology, the latter is only the backbone over which goods are traded, and the knowledge of the  volume of such trade is extremely important.
To be able to give predictions about the weight of connections, one needs to switch from an ensemble of binary graphs to one of weighted graphs.

The simplest weighted counterpart of the BCM is the WCM, which is a maximum-entropy ensemble of weighted networks where the constraint is the strength sequence, i.e. the total trade of each country in the case of the ITN.
Recent studies have shown that the higher-order binary quantities predicted by the WCM, as well as the corresponding weighted quantities, are very different from the observed counterparts \cite{Squartini2,Squartini3}. 
More specifically, the main limitation of the model is that of predicting a largely homogeneous and very dense (sometimes fully connected) topology.
Roughly speaking, the model excessively `dilutes' the total trade of each country by distributing it to almost all other countries. 
This failure in correctly replicating the  purely topological projection of the real network is the root of the bad agreement between expected and observed higher-order properties.

\subsubsection{Relation with the Gravity Model}
Just like the BCM has been related to the Fitness Model \cite{Diego1}, a variant of the WCM has been related to the Gravity Model \cite{Fronczak}.
The variant is actually a continuous version of the WCM, where the strength sequence is constrained and the weights are real numbers instead of integers.
When applied to the ITN, the model gives the following expectation for the weight of the links: 
\begin{equation}
 \langle w_{ij} \rangle =T\cdot g_i g_j	\qquad \forall i,j
\label{eq_ff}
\end{equation}
where $T$ is the total strength in the network, and $g_i$ is the re-scaled $GDP$ as before \cite{Fronczak}.
In essence, the above expression identifies again a relationship between the $GDP$ and the hidden variable (analogous to the fitness in the binary case) specifying the strength of a node.

Equation (\ref{eq_ff}) coincides with eq.(\ref{eq_gravity}) where $\beta=1$ and $\gamma=0$. 
The model therefore corresponds to a particularly simple version of the Gravity Model.
Indeed, the model reproduces reasonably well the observed non-zero weights of the ITN  \cite{Fronczak}.
However, just like the Gravity Model, the model predicts a complete graph where $a_{ij}=1~~ \forall {i,j}$, and dramatically fails in reproducing the binary architecture of the network.
This can be easily shown by realizing that the continuous nature of edge weights, which can take non-negative real values in the model, implies that there is a zero probability of generating zero weights (i.e. missing links).
We will show the prediction of this model in comparison with our results later on in the paper.

\section{A GDP-driven model of the ITN}
Motivated by the challenge to satisfactorily model both the topology and the weights of the ITN, the ECM has been recently proposed as an improved model of this network \cite{Squartini6}. 
The ECM focuses on weighted networks, and can enforce the degree and strength sequence simultaneously \cite{Squartini5}.
It builds on the so-called generalized Bose-Fermi distribution that was first introduced as a null model of networks with coupled binary and weighted constraints \cite{Garlaschelli2}.

In the ECM, the degree and strength sequence can be constrained by writing the following Hamiltonian:
\begin{equation}
H(\mathbf{W})=\sum_{i=1}^N \left[\alpha_i k_i (\mathbf{W})+\beta_is_i(\mathbf{W})\right]
\end{equation}
where the strength sequence is defined as $s_i(\mathbf{W})=\sum_{j\neq i}^N w_{ij},\:\forall\:i$ and the degree sequence as $k_i(\mathbf{W})=\sum_{j\neq i}^N a_{ij}=\sum_{j\neq i}^N \Theta[w_{ij}],\:\forall\:i$. As a result, the probability of a given configuration $\mathbf{W}$ can be written as

\begin{equation}
P(\mathbf{W})=
\frac{e^{-H(\mathbf{W})}}{\sum_{\mathbf{W}'}e^{-H(\mathbf{W}')}}
=\prod_{i<j} \frac{(x_ix_j)^{a_{ij}} (y_iy_j)^{w_{ij}}(1-y_iy_j)}{1-y_iy_j+x_ix_jy_iy_j}
\end{equation}
\noindent with $x_i\equiv e^{-\alpha_i}$ and $y_i\equiv e^{-\beta_i}$. 
The ECM gives the following predictions about the probability of a link ($\langle a_{ij} \rangle$) and the expected weight of the link ($\langle w_{ij} \rangle$):
\begin{eqnarray}
\langle a_{ij} \rangle&=&\frac{x_ix_jy_iy_j}{1-y_iy_j+x_ix_jy_iy_j}=p_{ij}\label{eq_pp}\\
\langle w_{ij} \rangle&=&\frac{x_ix_jy_iy_j}{(1-y_iy_j+x_ix_jy_iy_j)(1-y_iy_j)}=\frac{p_{ij}}{1-y_iy_j}.\label{eq_ww}
\end{eqnarray}

According to the maximum-likelihood method proposed in \cite{Squartini5}, the vectors of unknowns $\vec{x}$ and $\vec{y}$ can be numerically found by solving the system of $2N$ coupled equations
\begin{eqnarray}
\langle k_i(\mathbf{W})\rangle=\sum_{j\neq i}^N p_{ij}=k_i(\mathbf{W}^*)\qquad\forall\:i\label{eq_kk}\\
\langle s_i(\mathbf{W})\rangle=\sum_{j\neq i}^N \langle w_{ij}\rangle=s_i(\mathbf{W}^*)\qquad\forall\:i\label{eq_ss}
\end{eqnarray}
and will be indicated as $\vec{x}^*$ and $\vec{y}^*$. These unknown parameters can be treated as fitness parameters which control the probability of forming a link and the expected weight of that link simultaneously. 

The application of the ECM to various real-world networks shows that the model can accurately reproduce the higher-order empirical properties of these networks \cite{Squartini5}. 
When applied to the ITN in particular, the ECM replicates both binary and weighted empirical properties, for different levels of disaggregation, and for several years (temporal snapshots) \cite{Squartini6}.
Indeed, in fig. \ref{fig3} we show the higher-order binary quantities (average nearest neighbor degree and clustering coefficient) as well as their weighted ones (average nearest neighbor strength and weighted clustering coefficient) for the 2002 snapshot of the ITN. 
We compare the observed values (red points) and the corresponding quantities predicted by the ECM (blue points). 
The mathematical expressions for all these quantities are provided in the Appendix.
We find a very good agreement between data and model, confirming the recent results in \cite{Squartini6} for the data set we are using here.
We also confirmed that these results are robust for several temporal snapshots \cite{Squartini6}.

\subsection{A weighted fitness model of trade}
Considering the promising results of the ECM, we now make a step forward and check whether the hidden variables $x_i$ and $y_i$, which effectively reproduce the observed ITN, can be thought of as `fitness' parameters having a clear economic interpretation.  
This amounts to check whether the relation shown previously in fig. \ref{fig2} for the purely binary case can be generalized in order to find a macroeconomic interpretation to the abstract fitness parameters in the general weighted case as well.

\begin{figure}
\includegraphics[scale=0.6]{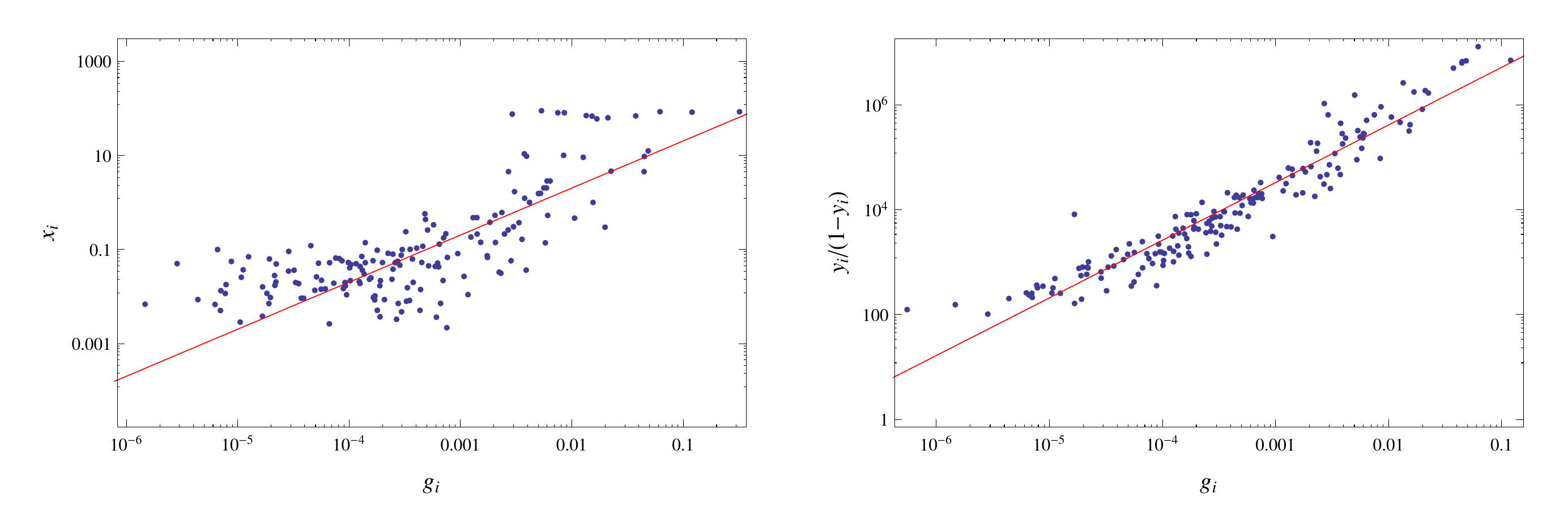}
\caption{\label{fig:epsart} The calculated $x_i$ (left panel)  and $\frac{y_i}{1-y_i}$ (right panel) compared with  the $g_i$ (rescaled GDP) for each country for the undirected binary aggregated ITN in the 2002 snapshot, with the linear fits (in log-log scale) $x_i= \sqrt{a}\cdot g_i$, and $\frac{y_i}{1-y_i}=b \cdot g_i^c$ (red lines), where a, b, and c are the fitted constant parameters per year.}
\label{fig4}
\end{figure}


In fig. \ref{fig4} we show the  relationship between the two parameters $x_i$ and $y_i$ and the rescaled $GDP$ ($g_i$) for each country of the ITN in the 2002 snapshot. 
We find strong correlations between these quantities.
The fitness parameter $x_i$ turns out to be in a roughly linear relation with the rescaled GDP $g_i$, fitted by the curve
\begin{equation}
x_i=\sqrt{a}\cdot g_i
\label{eq_xx}
\end{equation}
where $\sqrt{a}$ is the fitted constant, and $g_i=\frac{GDP_i}{\sum_i GDP_i}$ (all the $GDP$s are relative to that specific year).
It should be noted that this relation is similar to that found between $z_i$ and $g_i$ in the BCM and shown previously in fig.\ref{fig2}, but less accurate. This observation will be useful later.
By contrast, since the $GDP$ is an unbounded quantity, while the fitness parameter $y_i$ is bounded between 0 and 1 (this is a mathematical property of the model \cite{Garlaschelli2,Squartini5}), the relation between $y_i$ and $g_i$ is necessarily highly nonlinear.
A simple functional form for such a relationship is given by
\begin{equation}
y_i=\frac{b \cdot g_i^c}{1+b \cdot g_i^c}.
\label{eq_yy}
\end{equation}
Indeed, fig. \ref{fig4} confirms that the above expression provides a very good fit to the data.

We checked that the above results hold systematically over time, for each snapshot of the ITN in our data set.
This implies that, in a given year, we can insert eqs.(\ref{eq_xx}) and (\ref{eq_yy}) into eqs.(\ref{eq_pp}) and (\ref{eq_ww}) to obtain a GDP-driven model of the ITN structure for that year.
Such a model highlights that the $GDP$ has a crucial role in shaping both the binary and the weighted properties of the ITN.
While this was already expected on the basis of the aforementioned results obtained using the BCM and the WCM (or the corresponding simplified gravity model) separately, finding the appropriate way to explicitly combine these results into a unified description of the ITN has remained impossible so far.
Rather than exploring in more detail the predictions of the GDP-driven model in the form described above, we first make some considerations leading to a simplification of the model itself.

\subsection{Reduced two-step model}
At this point, it should be noted that we arrived at two seemingly conflicting results. 
We showed that both the BCM and ECM give a very good prediction for the binary topology of the network. However, eqs.(\ref{eq_pzz}) and (\ref{eq_pp}), which specify the connection probability $p_{ij}$ in the two models, are significantly different. The comparable performance of the BCM and the ECM at a binary level (see figs.\ref{fig1} and \ref{fig3}) makes us expect that, when the specific values $\vec{z}^*$ and $\vec{x}^*$ are inserted into eqs.(\ref{eq_pzz}) and (\ref{eq_pp}) respectively, the values of the connection probability become comparable in the two models, despite the different mathematical expressions.

\begin{figure}
\includegraphics[scale=0.6]{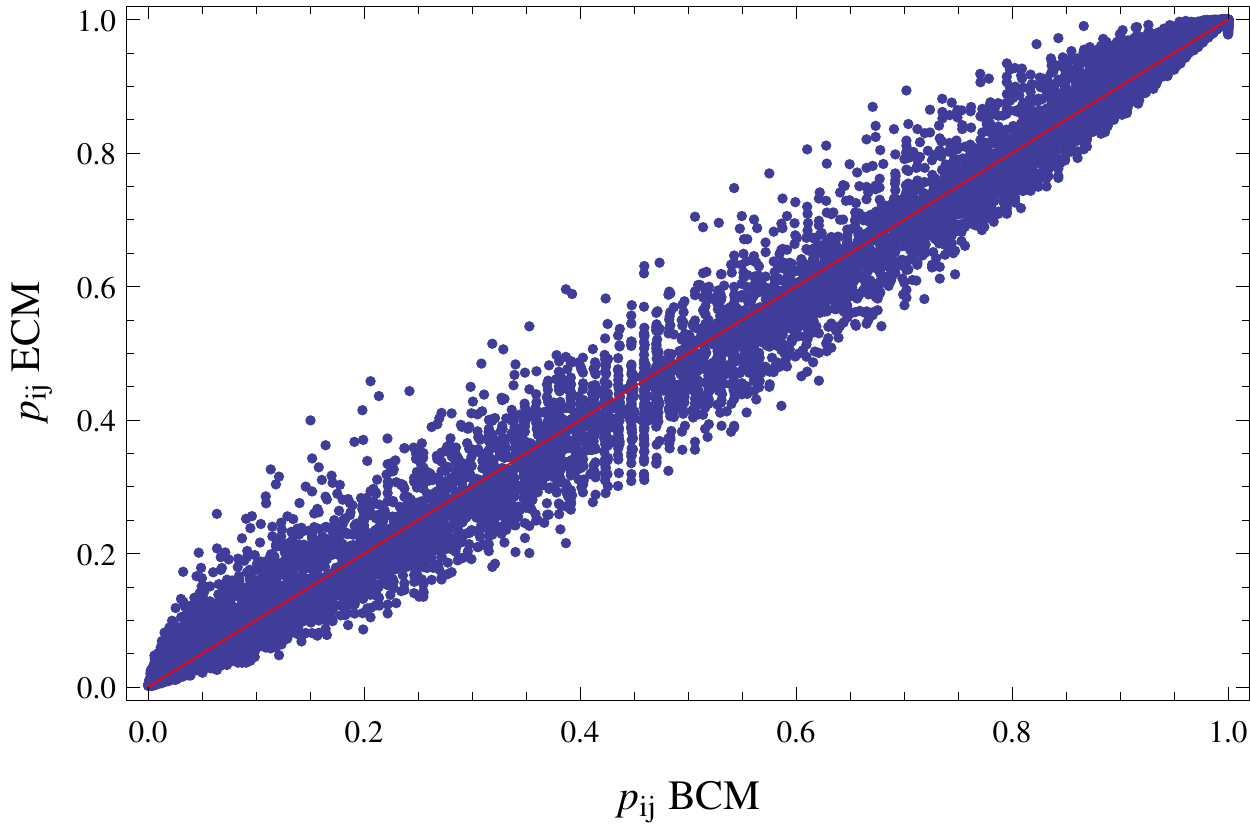}
\caption{\label{fig:epsart} The probability of forming a link in the ECM $p_{ij} ~ECM $ compared to the probability in the BCM $p_{ij}~ BCM$ for the undirected binary aggregated ITN in the 2002 snapshot. The red line describes the identity line. }
\label{fig5}
\end{figure}

In fig. \ref{fig5} we compare the the two probabilities for the ITN in the 2002 snapshot. 
Note that each point refers to the probability of creating a link between a pair of countries, which results in $\frac{N(N-1)}{2}$ points.
Indeed, we can see that the values are scattered along the identity line, confirming the expectation that the connection probability has similar value in the two models. 

The above result allows us to make a remarkable simplification.
In eqs.(\ref{eq_pp}) and (\ref{eq_ww}), we can replace the expression for $p_{ij}$ provided by the ECM with that provided by the BCM in eq.(\ref{eq_pzz}).
To avoid confusion, we denote the new probability with $p_{ij}^{ts}$, where $ts$ stands for `two-step', for a reason that will be clear immediately.
This results in the following equations for the expected network properties: 
\begin{eqnarray}
\langle a_{ij}\rangle^{ts}&=&\frac{z_iz_j}{1+z_iz_j}\equiv p_{ij}^{ts},\label{eq_pts}\\
\langle w_{ij}\rangle^{ts}&=&\frac{p_{ij}^{ts}}{1-y_iy_j}.\label{eq_wts}
\end{eqnarray}
where the $z_i$'s, and therefore the $p^{ts}_{ij}$'s, depend only on the degrees through eq.(\ref{MaxLike}), while the $y_i$'s and the $\langle w_{ij}\rangle^{ts}$'s depend on both strengths and degrees through eqs.(\ref{eq_kk}) and (\ref{eq_ss}).

In this simplified model the connection probability, which fully specifies the topology of the ensemble of networks, no longer depends on the strengths as in the ECM, while the weights still do. 
This implies that we can specify the model via a two-step procedure where we first solve the $N$ equations determining $p_{ij}^{ts}$ via the degrees, and then find the remaining variables determining $\langle w_{ij}\rangle^{ts}$ through the ECM.
For this reason, we denote the model as the Two-Step (TS) model.

The probability of a configuration $\mathbf{W}$ reads 
\begin{equation}
P^{ts}(\mathbf{W})=\prod_{i<j} q_{ij}^{ts}(w_{ij})
\end{equation}
where
\begin{equation}
q_{ij}^{ts}(w_{ij})=\frac{(z_iz_j)^{a_{ij}} (y_iy_j)^{w_{ij}-a_{ij}}(1-y_iy_j)^{a_{ij}}}{1+z_iz_j}
\label{prob}
\end{equation}
is the probability that a link of weight $w_{ij}$ connects the nodes (countries) $i$ and $j$.
The above probability has the same general expression as in the original ECM \cite{Squartini5}, but here $z_i$ comes from the estimation of the simpler BCM.
It is instructive to rewrite (\ref{prob}) as 
\begin{eqnarray}
q_{ij}^{ts}(0)&=&\frac{1}{1+z_iz_j}=(1-p_{ij}^{ts});\\
q_{ij}^{ts}(w)&=&p_{ij}^{ts}(y_iy_j)^{w-1}(1-y_iy_j),\:\forall\:w>0
\label{prob2}
\end{eqnarray}
to highlight the random processes creating each link. 
As a first step, one determines whether a link is created or not with a probability $p_{ij}^{ts}$. If a link (of unit weight) is indeed established, a second attempt determines whether the weight of the same link is increased by another unit (with probability $y_iy_j$)
or whether the process stops (with probability $1-y_iy_j$).
Iterating this procedure, the probability that an edge with weight $w$ is established between nodes $i$ and $j$ is given precisely by $q_{ij}^{ts}(w)$ in eq.(\ref{prob2}).
The expected weight $\langle w_{ij}\rangle^{ts}$ is then correctly retrieved as $\sum_{w=0}^{+\infty}w\cdot q_{ij}^{ts}(w)$.

Using the relations found in eqs.(\ref{eq_zz}) and (\ref{eq_yy}), we can input the $g_i$ as the fitness parameters into eqs.(\ref{eq_pts}) and (\ref{eq_wts}) to get the following expressions that mathematically characterize our GDP-driven specification of the TS model: 
\begin{eqnarray} \label{Twostep1}
\langle a_{ij}\rangle^{ts}&=&\frac{a g_ig_j}{1+a g_ig_j}\equiv p_{ij}^{ts}\\ \label{Twostep2}
\langle w_{ij}\rangle^{ts}&=&p_{ij}^{ts} \frac{(1+b g_i^c)(1+b g_j^c)}{(1+b g_i^c+b g_j^c)}.
\end{eqnarray}
The above equations can be used to reverse the approach used so far: rather than using the $2N$ free parameters of the ECM ($\vec{x}$ and $\vec{y}$) or of the TS model ($\vec{z}$  and $\vec{y}$) to fit the models on the observed values of the degrees and strengths, we can now use the knowledge of the GDP of all countries to obtain a model that only depends on the three parameters $a$, $b$, $c$.
Assigning values to these parameters  can be done using two techniques: maximization of the likelihood function and non-linear curve fitting. 
Since the model is a two-step one, we can first assign a value to the parameter $a$, and only in the second step (once $a$ is set) we fit the parameters $b$ and $c$.

We chose to fix $a$ by maximizing the likelihood function \cite{Diego2}, which results in  constraining the expected number of links to the observed number ($\langle L\rangle=L$), as in \cite{Diego1}.
Fixing the values of $b$ and $c$ is slightly more complicated. 
Since the model uses the approximated expressions of the TS model, rather than those of the original ECM model, maximizing the likelihood function in the second step no longer yields the desired condition $\langle T\rangle=T$, where $T$ is the total strength in the network.
Similarly, extracting the parameters from the fit as shown in fig. \ref{fig4} does not maintain the total strength in the network.  
In absence of any a-priori preference, we chose the latter procedure, due to its relative numerical simplicity with respect to the former one.

\begin{figure*}{}
\includegraphics[scale=0.6]{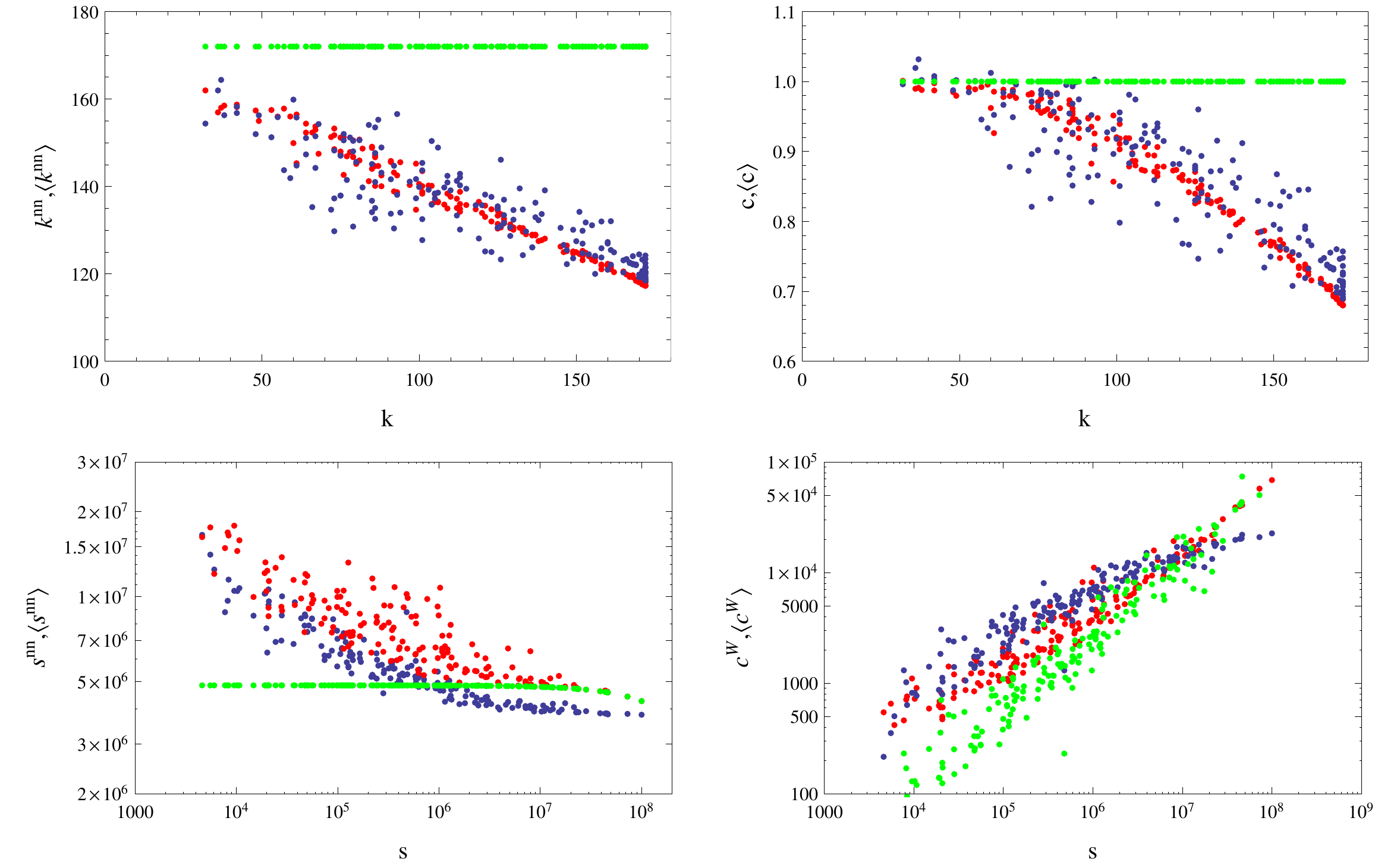}
\caption{\label{fig:epsart}Comparison between the observed properties (red points), the corresponding ensemble averages of the $GDP$-driven two-steps model (blue points) and the $GDP$-driven WCM model (green points), of the aggregated ITN in 2002.
Top left: Average Nearest Neighbor Degree $k^{nn}_i$ versus degree $k_i$. Top right: Binary Clustering Coefficient $c_i$ versus degree $k_i$. Bottom left: Average Nearest Neighbor Strength $s^{nn}_i$ versus strength $s_i$. Bottom right: Weighted Clustering Coefficient $c^{w}_i$ versus strength $s_i$.}
\label{fig6}
\end{figure*}

In fig. \ref{fig6} we show a comparison between the higher-order observed properties of the ITN in 2002 and their expected counterparts predicted by the GDP-driven TS model.
Again, the mathematical expressions of these properties are provided in the Appendix.
As a baseline comparison, we also show the predictions of the $GDP$-driven WCM model with continuous weights described by eq.(\ref{eq_ff}) \cite{Fronczak}, which coincides with a simplified version of the gravity model as we mentioned. 

We see that the GDP-driven TS model reproduces the empirical trends very well. Of course, as expected, the predictions in fig.\ref{fig6} (which use only three free parameters) are more noisy than those in fig.\ref{fig3} (which use $2N$ free parameters). This is due to the fact that eqs.(\ref{eq_zz}) and (\ref{eq_yy}) describe fitting curves rather than exact relationships.
Importantly, our model performs significantly better than the WCM/gravity model in replicating both binary and weighted properties.
Again, the drawback of these models lies in the fact that they predict a fully connected topology and a relatively homogeneous network.

It should also be noted that the plot of average nearest neighbor strength ($s^{nn}$) predicted by our model is slightly shifted with respect to the observed points. This effect is due to the fact that, as we mentioned, the total strength $T$ (hence the average trend of the $s^{nn}$) is only approximately reproduced by our model, as a result of the simplification from the ECM to the TS model.

As for all the other results in this paper, we checked that our findings are robust over the entire time span of our data set.
We therefore conclude that the ECM model, as well as its simplified TS variant, can be successfully turned into a fully GDP-driven model that simultaneously reproduces both the topology and the weights of the ITN.\\

The success of the TS model has an important interpretation. Looking back at eqs.(\ref{eq_pts}) and (\ref{Twostep1}), we recall that the effect of the TS approximation is the fact that the connection probability $p_{ij}^{ts}$ can be estimated separately from the weights $\langle w_{ij}\rangle^{ts}$, using only the knowledge of the degree sequence if eq.(\ref{eq_pts}) is used, or the GDP and total number of links if eq.(\ref{Twostep1}) is used, while discarding that of the strengths.
By contrast, the estimation of the expected weights $\langle w_{ij}\rangle^{ts}$ cannot be carried out separately, as it requires that the connection probability $p_{ij}^{ts}$ appearing in eqs.(\ref{eq_wts}) and (\ref{Twostep2}) is estimated first.
This asymmetry of the model means that the topology of the ITN can be successfully inferred without any information about the weighted properties, while the weighted structure cannot be inferred without topological information.
The expressions defining the TS model provide a mathematical explanation for this otherwise puzzling effect that has already been documented in previous analyses of the ITN \cite{Squartini1,Squartini2,Squartini3,Squartini6}.

\section{Conclusions}
In this paper we have introduced a novel $GDP$-driven model which successfully reproduces both the binary and weighted properties of the ITN.
The model uses the $GDP$ of countries as a sort of macroeconomic fitness, and reveals the existence of strong relations between the $GDP$ and the model parameters controlling the formation and the volume of trade relations.  
In the light of the limitations of the existing models (most notably the  binary-only nature of the fitness model and the weighted-only nature of the gravity model), we believe that our results represent a promising step forward in the development of a unified model of the ITN structure. 
We have also shown that the full ECM model can be effectively reduced to the simpler TS model. The success of the latter provides a mathematical explanation of an otherwise puzzling asymmetry, namely the fact that purely topological properties can be successfully predicted without knowing the weights, while weighted network properties can only be predicted if the topological ones are preliminary estimated.
Future work should explore how to further improve these results and possibly expand them by introducing additional macroeconomic parameters like geographic distance, thus fully bridging the gap between network-based and gravity-based approaches to the structure of international trade.

\acknowledgements
AA and DG acknowledge support from the Dutch Econophysics Foundation (Stichting Econophysics, Leiden, the Netherlands) with funds from beneficiaries of Duyfken Trading Knowledge BV, Amsterdam, the Netherlands. 
This work was also supported by the EU project MULTIPLEX (contract 317532) and the Netherlands Organization for Scientific Research (NWO/OCW).

\appendix
\section{Higher-Order Properties}
In this appendix we give a summarized description of the binary and weighted network quantities which are studied in this paper.
Specifically, we first show how the properties are measured over a real network, and then how the expected values under the ECM and the TS model are constructed.

\subsection{Observed Properties}

Let us note a weighted undirected network as a square matrix $\mathbf{W}$, where  the specific entry $w_{ij}$ represents the edge weight between country $i$ and country $j$. 
The binary representation of the network is noted by a binary matrix $\mathbf{A}$, where  the entries are $a_{ij}= \Theta[w_{ij}],\:\forall\:{i,j}$.\\

We compute the Average Nearest Neighbor Degree as:
\begin{equation}
 k_i^{nn}(\mathbf{W})=\sum_{j\neq i} \frac{a_{ij} k_j}{k_i}=\frac{\sum_{j\neq i}\sum_{k\neq j}a_{ij}a_{jk}}{\sum_{j\neq i}a_{ij}}.
\end{equation}
Its calculated as the arithmetic mean of the degrees of the neighbors of a specific node, which is a measure of correlation between
the degrees of adjacent nodes.\\

The Binary Clustering Coefficient has the following expression:
\begin{equation}
 c_i(\mathbf{W})=\frac{\sum_{j\neq i}\sum_{k\neq i,j}a_{ij}a_{jk}a_{ki}}{\sum_{j\neq i}\sum_{k\neq i,j}a_{ij}a_{ki}}.
\end{equation}
It is a measure of the tendency to which nodes in a graph form cluster together. 
More specifically, it counts how many closed triangles are attached to each node with respect to all the possible triangles.\\

The corresponding weighted properties are the Average Nearest Neighbor Strength and the weighted Clustering Coefficient.
The Average Nearest Neighbor Strength, defined as:
\begin{equation}
 s_i^{nn}(\mathbf{W})=\sum_{j\neq i} \frac{a_{ij} s_j}{k_i}=\frac{\sum_{j\neq i}\sum_{k\neq j}a_{ij}w_{jk}}{\sum_{j\neq i}a_{ij}}
\end{equation}
where $s_i=\sum_j w_{ij}$ is the strength (total flow) of a country.
The $s_i^{nn}$ measure the average strength of the neighbors for a specific node $i$. 
Like its binary counterpart, it gives the magnitude of activity of a specific node neighbors (weighted activity). 

The weighted Clustering Coefficient \cite{Giorgio_clustering} defined as: 
\begin{equation}
 c_i^{W}(\mathbf{W})=\frac{\sum_{j\neq i}\sum_{k\neq i,j}(w_{ij}w_{jk}w_{ki})^{\frac{1}{3}}}{\sum_{j\neq i}\sum_{k\neq i,j}a_{ij}a_{ki}}.
\label{cluster}
 \end{equation}
The $c_i^{W}(\mathbf{W})$ is a measure of the weight density in the neighborhood of a node.
It classify the tendency of a specific node to cluster in a triangle taking into account also the edge-values.\\

Now, the measured properties of the real network need to be compared with the reproduced properties of the different models. 
These reproduced properties are the expected values of the maximum entropy ensemble that each model id generating, and can be calculated analytically. 
The expected values can be obtained by simply replacing $a_ij$ with the probability $p_{ij}$ for the different models ($p_{ij}$ is different to each model). This next step is what we will discuss in the next sections.

\subsection{Expected values in the BCM and ECM}

Since the BCM model is only dealing with the binary representation, we will have expected values just for the two binary higher-order properties. While the ECM gives expectations for the weighted counterparts of the binary properties.\\

For the binary higher-order properties, we replace $a_ij$ with $p_{ij}$ which is the probability of creating a link, and also the expected value of the edge $p_{ij}=\langle a_{ij} \rangle$. This simple procedure yields the analytic formula of the expected value for the properties.  
We compute the expected Average Nearest Neighbor Degree as:
\begin{equation}
 \langle k_i^{nn}\rangle =\frac{\sum_{j\neq i}\sum_{k\neq j}p_{ij}p_{jk}}{\sum_{j\neq i}p_{ij}}
\end{equation}

and the expected Binary Clustering Coefficient as:
\begin{equation}
\langle c_i\rangle=\frac{\sum_{j\neq i}\sum_{k\neq i,j}p_{ij}p_{jk}p_{ki}}{\sum_{j\neq i}\sum_{k\neq i,j}p_{ij}p_{ki}}
\end{equation}
where for the BCM model we input $p_{ij}=\frac{z_i z_j}{1+z_iz_j}$, and for the ECM the more complex term $p_{ij}=\frac{x_i x_jy_i y_j}{1-y_iy_j+x_i x_jy_i y_j}$

In the weighted case (weighted higher-order properties), we are left only with the ECM. 
The expected Average Nearest Neighbor Strength is calculated  as:
\begin{equation}
\langle s_i^{nn}\rangle=\frac{\sum_{j\neq i}\sum_{k\neq j}p_{ij}\langle w_{jk}\rangle}{\sum_{j\neq i}p_{ij}}
\end{equation}
where $\langle w_{jk}\rangle=\frac{x_i x_jy_i y_j}{(1-y_iy_j+x_i x_jy_i y_j)(1-y_iy_j)}$ and we input the $p_{ij}$ of the ECM model as before.\\

In the expected value of the $c^{W}$ we should be more careful, since it is necessary to calculate the expected product of (powers of) distinct matrix entries 
\begin{equation}
\langle c_i^{W}\rangle=\frac{\sum_{j\neq i}\sum_{k\neq i,j}\langle( w_{ij} w_{jk} w_{ki})^{\frac{1}{3}}\rangle}{\sum_{j\neq i}\sum_{k\neq i,j}p_{ij}p_{ki}}.
\label{cluster}
 \end{equation}

We know that
\begin{equation}
 \langle \sum_{i\neq j\neq k} w_{ij}^{\alpha} \cdot w_{jk}^{\beta}\cdot ... \rangle= \sum_{i\neq j\neq k} \langle w_{ij}^{\alpha}\rangle \cdot  \langle w_{jk}^{\beta}\rangle \cdot \langle ... \rangle
\end{equation}
with the generic term for the ECM case
\begin{equation}
 \langle w_{ij}^{\gamma}\rangle=\sum_{0}^{\infty} w^{\gamma} q_{ij} ( w | \vec{x}, \vec{y})= \frac{x_ix_j(1-y_iy_j)Li_{\gamma}(y_iy_j)}{1-y_iy_j+x_i x_jy_i y_j}
\end{equation}
where $Li_n(R)=\sum_{l=1}^{\infty} \frac{R^l}{l^n}$ is the $n$th polylogarithm of $R$.
For a more comprehensive description please refer to \cite{Squartini5}.

\subsection{Expected values in the TS model}
Here again we use the known expressions for the properties and replacing the terms $p^{ts}_{ij}$ and $w^{ts}_{ij}$ with the expected values $\langle a_{ij}\rangle$ and $\langle w_{ij} \rangle$ correspondingly.
However, here the expected values are a function of the $GDP$ of the countries, or more specifically the re-scaled $GDP$ $g_i$. 
The expresions for the higher-order binary properties are as before : 
\begin{equation}
 \langle k_i^{nn}\rangle =\frac{\sum_{j\neq i}\sum_{k\neq j}p^{ts}_{ij}p^{ts}_{jk}}{\sum_{j\neq i}p^{ts}_{ij}}
\end{equation}
and
\begin{equation}
\langle c_i\rangle=\frac{\sum_{j\neq i}\sum_{k\neq i,j}p^{ts}_{ij}p^{ts}_{jk}p^{ts}_{ki}}{\sum_{j\neq i}\sum_{k\neq i,j}p^{ts}_{ij}p^{ts}_{ki}}
\end{equation}
where $p^{ts}_{ij}=\frac{a g_ig_j}{1+a g_ig_j}$.

In the weighted case,the expected Average Nearest Neighbor Strength is calculated  as:
\begin{equation}
\langle s_i^{nn}\rangle=\frac{\sum_{j\neq i}\sum_{k\neq j}p^{ts}_{ij}\langle w^{ts}_{jk}\rangle}{\sum_{j\neq i}p^{ts}_{ij}}
\end{equation}
where $\langle w_{ij}\rangle^{ts}=\frac{a g_ig_j}{1+a g_ig_j}\cdot \frac{(1+b g_i^c)(1+b g_j^c)}{(1+b g_i^c+b g_j^c)}$ and we input the $p^{ts}_{ij}$ of the Two-Step model as before.

For convenience reasons we will write the expression for the weighted Clustering Coefficient $c^{W}$ first as a function of the fitness parameters $z_i$ and $y_i$, and later replaced them with the corresponding $GDP$ terms. 
The expected the expected value of the $c^{W}$ in the Two-Step case is : 
\begin{equation}
\langle c_i^{W}\rangle=\frac{\sum_{j\neq i}\sum_{k\neq i,j}\langle( w_{ij} w_{jk} w_{ki})^{\frac{1}{3}}\rangle^{ts}}{\sum_{j\neq i}\sum_{k\neq i,j}p^{ts}_{ij}p^{ts}_{ki}}.
\label{cluster}
 \end{equation} 
As before we observe that 
\begin{equation}
 \langle \sum_{i\neq j\neq k} w_{ij}^{\alpha} \cdot w_{jk}^{\beta}\cdot ... \rangle= \sum_{i\neq j\neq k} \langle w_{ij}^{\alpha}\rangle \cdot  \langle w_{jk}^{\beta}\rangle \cdot \langle ... \rangle
\end{equation}

with the generic term for the Two-Step model
\begin{equation}
 \langle w_{ij}^{\gamma}\rangle=\sum_{0}^{\infty} w^{\gamma} q_{ij} ( w | \vec{z}, \vec{y})= \frac{z_iz_j(1-y_iy_j)Li_{\gamma}(y_iy_j)}{(1+z_iz_j)y_iy_j}
 \label{poly}
\end{equation}
where $Li_n(R)=\sum_{l=1}^{\infty} \frac{R^l}{l^n}$ is the $n$th polylogarithm of $R$.

Once we input the expressions of $z_i$ and $y_i$
\begin{eqnarray}
z_i&=& \sqrt{a}\cdot g_i,\nonumber\\
y_i&=&\frac{b \cdot g_i^c}{1+b \cdot g_i^c}
\label{eqsfit}
\end{eqnarray}
equation (\ref{poly}) yields
\begin{equation}
 \langle w_{ij}^{\gamma}\rangle= \frac{a g_ig_j}{1+a g_ig_j} \cdot \frac{1+cg_i^c+bg_j^c}{b^2g_i^cg_j^c}Li_{\gamma}\left(\frac{b^2 g_i^c g_j^c}{(1+bg_i^c)(1+bg_j^c)}\right)
\end{equation}


\begin{thebibliography}{99}
\bibitem{economicnetworks}
F. Schweitzer, G. Fagiolo, D. Sornette, F. Vega-Redondo, A. Vespignani, D.R. White. {\it Economic networks: The new challenges}, Science {\bf 325(5939)}, 422 (2009).

\bibitem{Kali1} 
R. Kali and J. Reyes,{\it The architecture of globalization: a network approach to international economic integration},
Journal of International Business Studies {\bf38}, 595 (2007).

\bibitem{Kali2} 
R. Kali and J. Reyes, {\it Financial contagion on the international trade network},
Economic Inquiry {\bf 48}, 1072 (2010).

\bibitem{integration}
S. Schiavo, J. Reyes, G. Fagiolo, {\it International trade and financial integration: a weighted network analysis}, Quantitative Finance {\bf 10(4)}, 389-399 (2010).

\bibitem{Serrano} 
A. Serrano and M. Boguna,{\it Topology of the world trade web},
Phys. Rev. E {\bf68},015101(R) (2003).

\bibitem{Diego1}
D. Garlaschelli, M. I. Loffredo, {\it Fitness-dependent topological properties of the World Trade Web}, 
Phys. Rev. Lett. {\bf 93}, 188701 (2004).

\bibitem{Garlaschelli} 
D. Garlaschelli and M. Loffredo, {\it Structure and Evolution of the World Trade Network}, 
Physica A {\bf355}, 138 (2005).

\bibitem{Vespignani}
A. Serrano, M. Boguna, and A. Vespignani,{\it Patterns of dominant flows in the world trade web},
J. Econ. Interact. Coord.{\bf 2}, 111 (2007).

\bibitem{Caldarelli}
D. Garlaschelli, T. Di Matteo, T. Aste, G. Caldarelli, and M. Loffredo, {\it Interplay between topology and dynamics in the World Trade Web},
Eur. Phys. J. B {\bf 57}, 1434 (2007).

\bibitem{Fagiolo1}
G. Fagiolo, J. Reyes, S. Schiavo, {\it On the topological properties of the world trade web: A weighted network analysis}, Physica A {\bf 387(15)}, 3868-3873 (2008).

\bibitem{Fagiolo2}
G. Fagiolo, J. Reyes, S. Schiavo, {\it World-trade web: Topological properties, dynamics, and evolution}, Physical Review E {\bf 79(3)}, 036115 (2009).

\bibitem{Fagiolo3}
G. Fagiolo, J. Reyes, S. Schiavo, {\it The evolution of the world trade web: a weighted-network analysis}, Journal of Evolutionary Economics, {\bf 20(4)}, 479-514 (2010).

\bibitem{Barigozzi} 
M. Barigozzi, G. Fagiolo, and D. Garlaschelli, {\it Multinetwork of international trade: A commodity-specific analysis},
Physical Review E {\bf 81}, 046104 (2010).


\bibitem{DeBene1}
L. De Benedictis, L. Tajoli, {\it The world trade network}, The World Economy {\it 34(8)}, 1417-1454 (2011).

\bibitem{Squartini1}
T. Squartini, G. Fagiolo, D. Garlaschelli, {\it Randomizing world trade. I. A binary network analysis},
Phys. Rev. E {\bf 84}, 046117 (2011).

\bibitem{Squartini2}
T. Squartini, G. Fagiolo, D. Garlaschelli, {\it Randomizing world trade. II. A weighted network analysis}, 
Phys. Rev. E {\bf 84}, 046118 (2011).

\bibitem{Squartini3}
G. Fagiolo, T. Squartini, D. Garlaschelli, {\it Null Models of Economic Networks: The Case of the World Trade Web}, 
J. Econ. Interac. Coord. {\bf 8}(1), 75-107 (2013).

\bibitem{Fronczak}
A. Fronczak, P. Fronczak, J.A. Holyst {\it Statistical mechanics of the international trade network},
Phys. Rev. E  {\bf 85}, 056113 (2012).


\bibitem{Squartini7}
T. Squartini and D. Garlaschelli, {\it In Self-Organizing Systems}, Lecture Notes in Computer Science 7166, pp. 24–35 (Springer, 2012).


\bibitem{Pietronero}
M. Cristelli, A. Gabrielli, A. Tacchella, G. Caldarelli, L. Pietronero, {\it Measuring the Intangibles: A Metrics for the Economic Complexity of Countries and Products},
PLoS ONE {\bf 8} 0070726 (2013).

\bibitem{Tinbergen1}
J. Tinbergen, {\it Shaping the World Economy: Suggestions for an International Economic Policy}, 
(The Twentieth Century Fund, New York, 1962).

\bibitem{Tinbergen2}
T. Squartini and D. Garlaschelli, {\it Jan Tinbergen's legacy for economic networks: from the gravity model to quantum statistics}, in Econophysics of Agent-Based Models, pp. 161-186 (Springer International Publishing, 2014).

\bibitem{Giorgio_gravity0}
G. Fagiolo, {\it The international-trade network: gravity equations and topological properties}, Journal of Economic Interaction and Coordination {\bf 5(1)}, 1-25 (2010).

\bibitem{Giorgio_gravity}
M. Duenas, G. Fagiolo, {\it Modeling the International-Trade Network: A Gravity Approach},
Journal Of Economic Interaction And Coordination {\bf 8} 155-178 (2013).

\bibitem{DeBene2}
L. De Benedictis, D. Taglioni, {\it The gravity model in international trade}, in The Trade Impact of European Union Preferential Policies, pp. 55-89 (Springer Berlin Heidelberg, 2011).

\bibitem{myreciprocity}
D. Garlaschelli, M.I. Loffredo, {\it Patterns of link reciprocity in directed networks}, Physical Review Letters {\bf 93(26)}, 268701 (2004).

\bibitem{Giorgio_symmetry}
G. Fagiolo, {\it Directed or Undirected? A New Index to Check for Directionality of Relations in Socio-Economic Networks}, Economics Bulletin {\bf 3(34)}, 1-12 (2006).

\bibitem{mywreciprocity}
T. Squartini, F. Picciolo, F. Ruzzenenti, D. Garlaschelli, {\it Reciprocity of weighted networks}, Scientific Reports, 3 (2013).

\bibitem{Wells}
Wells, S.{\it Financial interlinkages in the United Kingdom’s interbank market and the risk of contagion},
Bank of England Working Paper, No. 230/2004  (2004).

\bibitem{Bargigli}
Bargigli, L., and Gallegati, M. {\it  Random digraphs with given expected degree sequences: A model for economic networks},
Journal of Economic Behavior \& Organization, {\bf 78}, 396-411 (2011).

\bibitem{Musmeci}
Musmeci, N., Battiston, S., Caldarelli, G., Puliga, M., Gabrielli, A. {\it Bootstrapping topological properties and systemic risk of complex networks using the fitness model}, 
Journal of Statistical Physics, {\bf 151}, 720-734 (2013).

\bibitem{Caldarelli 2}
Caldarelli, G., Chessa, A., Pammolli, F., Gabrielli, A., and Puliga, M {\it Reconstructing a credit network}
Nature Physics, {\bf9}, 125-126 (2013).

\bibitem{Squartini5}
R. Mastrandrea, T. Squartini, G. Fagiolo, D. Garlaschelli, {\it Enhanced reconstruction of weighted networks from strengths and degrees},
New J. Phys. {\bf 16}, 043022 (2014).

\bibitem{MaxEntropy}
T. Squartini, D. Garlaschelli, {\it Analytical maximum-likelihood method to detect patterns in real networks}
New J. Phys. {\bf 13}, 083001 (2011).

\bibitem{Garlaschelli2}
D. Garlaschelli, M. I. Loffredo {\it Generalized Bose-Fermi Statistics and Structural Correlations in Weighted Networks},
Phys. Rev. Lett. {\bf102}, 038701 (2009).

\bibitem{Squartini6}
R. Mastrandrea, T. Squartini, G. Fagiolo, D. Garlaschelli, {\it Intensive and extensive biases in economic networks: reconstructing the world trade multiplex}, 
arXiv:1402.4171, (2014).




\bibitem{Diego2}
D. Garlaschelli, M.I. Loffredo, {\it Maximum likelihood: Extracting unbiased information from complex networks}, Physical Review E {\bf 78(1)}, 015101 (2008).

\bibitem{Squartini4}
F. Picciolo, T. Squartini, F. Ruzzenenti, R. Basosi, D. Garlaschelli, {\it The role of distances in the World Trade Web}, Proceedings of the Eighth International Conference on Signal-Image Technology \& Internet-Based Systems (SITIS 2012), pp. 784-792 (edited by IEEE) (2013).



\bibitem{comtrade}
http://comtrade.un.org/.

\bibitem{Gleditsch}
K. S. Gleditsch,  {\it Expanded trade and GDP data}, 
Journal of Conflict Resolution {\bf46} 712–24 (2002).

\bibitem{fitness}
G. Caldarelli, A. Capocci, P. De Los Rios, M. A. Mu\~{n}oz, {\it Scale-free networks from varying vertex intrinsic fitness}, Physical Review Letters, {\bf 89}(25), 258702  (2002).

\bibitem{Giorgio_clustering}
G. Fagiolo, {\it Clustering in complex directed networks},
Phys. Rev. E {\bf76}, 026107 (2007).



\end{thebibliography}
\end{document}